\documentclass[prb,reprint,amsmath,amssymb,floatfix,citeautoscript,superscriptaddress]{revtex4-1}
\usepackage{cancel}
\usepackage{amsmath}
\usepackage{graphicx}
\usepackage{amssymb}
\usepackage{amsthm}
\usepackage{bm}
\usepackage{dcolumn}
\usepackage{physics}
\usepackage{longtable}

\usepackage{txfonts}

\usepackage{color}
\usepackage[usenames,dvipsnames]{xcolor}
\definecolor{myblue}{rgb}{0,0,1}
\usepackage[breaklinks=true,colorlinks=true,linkcolor=myblue,urlcolor=myblue,citecolor=myblue]{hyperref}

\let\vr\undefined
\newcommand{\vr}{{\bm{r}}}

\newcommand{\vk}{{\bm{k}}}
\newcommand{\vp}{{\bm{p}}}

\newcommand{\vT}{{\bm{T}}}

\begin{document}

\title{
Thickness-dependent optical properties of layered hybrid organic-inorganic
halide perovskites:\\A tight-binding GW-BSE study
}

\author{Yeongsu Cho}
\affiliation{Department of Chemistry and James Franck Institute,
University of Chicago, Chicago, Illinois 60637, USA}
\author{Timothy C. Berkelbach}
\affiliation{Department of Chemistry, 
Columbia University, New York, New York 10027, USA}
\affiliation{Center for Computational Quantum Physics, Flatiron Institute, New York, New York 10010, USA}
\email{tim.berkelbach@gmail.com}

\begin{abstract}
We present a many-body calculation of the band structure and
optical spectrum of the layered hybrid organic-inorganic halide perovskites in
the Ruddlesden-Popper phase with the general formula
A$^\prime_2$A$_{n-1}$M$_n$X$_{3n+1}$, focusing specifically on the lead iodide
family.  We calculate the mean-field band structure with spin-orbit coupling,
quasiparticle corrections within the GW approximation, and optical spectra using
the Bethe-Salpeter equation.  The model is parameterized by first-principles
calculations and classical electrostatic screening, enabling an accurate but
cost-effective study of large unit cells and corresponding thickness-dependent
properties.  A transition of the electronic and optical properties from
quasi-two-dimensional behavior to three-dimensional behavior is shown for
increasing $n$ and the nonhydrogenic character of the excitonic Rydberg series
is analyzed.  The thickness-dependent 1s and 2s exciton energy levels are in
good agreement with recently reported experiments and the 1s exciton binding
energy is calculated to be 302 meV for $n=1$, 97~meV for $n=5$, and 37~meV for
$n=\infty$ (bulk MAPbI$_3$).
\end{abstract}

\maketitle


Hybrid organic-inorganic perovskites (HOIPs) are promising photovoltaic
materials, most recently showing a high power conversion efficiency of over 24\%
\cite{nrel}. A three dimensional bulk HOIP AMX$_3$ can be transformed into a
layered HOIP in the Ruddlesden-Popper phase A$^\prime_2$A$_{n-1}$M$_n$X$_{3n+1}$
by substituting a small organic cation A$^{+}$ by a bulkier one A$^{\prime+}$.
Common choices for the small organic cation are A$^+$=CH$_3$NH$_3^+$, NH$_4^+$;
for the bulkier cation are A$^{\prime+}$=C$_4$H$_9$NH$_3^+$,
C$_6$H$_5$C$_2$H$_4$NH$_3^+$; for the metal are M$^{2+}$=Sn$^{2+}$, Pb$^{2+}$;
and for the halide are X$^-$=Cl$^-$, I$^-$, Br$^-$.  A major drawback of the 3D
HOIPs for photovoltaics is their relatively fast degradation when exposed to
air, moisture, and light; in contrast, the layered HOIPs are more stable while
maintaining high power conversion efficiencies under working conditions
\cite{cao20152d, tsai2016high}.  The optical properties of layered HOIPs can
also be easily controlled by composition \cite{mcmeekin2016mixed}, enhancing
their flexibility for a variety of optoelectronic applications. Unlike the van
der Waals materials -- a prototypical family of layered materials including
graphene, hexagonal boron nitride, and the transition-metal dichalcogenides --
the layered HOIPs have sublayers that are covalently bonded.  This distinct
property makes the layered HOIPs an insightful mixed-dimensional platform for
investigating the transition of optoelectronic properties from two dimensional
to three dimensional.

The $n$-dependent properties of layered HOIPs have been experimentally
investigated, especially during the last five years~\cite{tanaka2003bandgap,
wu2015excitonic, stoumpos2016ruddlesden, stoumpos2017high, blancon2018scaling},
including mechanically exfoliated thin sheets of a layered HOIP
\cite{yaffe2015excitons, blancon2017extremely}. Early theoretical investigations
of the optical properties of layered HOIPs were performed using the effective
mass approximation, giving good estimates for the exciton binding energy and a
qualitative explanation of the essential physics~\cite{hong1992dielectric,
muljarov1995excitons, tanaka2003bandgap, tanaka2005image}. For a quantitative
analysis, \textit{ab initio} approaches such as density functional theory (DFT)
have been applied to study some $n$-dependent electronic
properties\cite{bala2018atomic} and were employed to build a semi-empirical
model of an exciton incorporating heterogeneous dielectric screening
\cite{blancon2018scaling}.  The GW approximation combined with the
Bethe-Salpeter equation (BSE) represents a standard many-body approach for the
accurate determination of the band structure and optical properties of
semiconductors.  While the 3D HOIPs have been extensively studied with the GW
approximation \cite{brivio2014relativistic, filip2014g, umari2014relativistic,
filip2015gw, mosconi2016electronic}, the analogous calculation for the layered
HOIPs is formidable due to the large number of atoms in the unit cell.
Recently, by replacing the organic A$^{\prime+}$ by a simpler cation, Cs$^+$, a
GW-BSE calculation has been performed on cubic
Cs$_2$PbI$_4$\cite{molina2018excitonic}, corresponding to the $n=1$
all-inorganic layered lead-halide perovskite.  For an overview of recent
progress on low dimensional hybrid perovskites, including their structural and
electronic properties, we refer to the excellent review by Katan, Mercier, and
Even~\cite{katan2019quantum}.

In this work, we pursue a semiempirical GW-BSE approach in order to study the
large unit cells encountered in layered HOIPs.  Our approach differs from
previous theoretical works because it is almost entirely parameterized by
first-principles calculations and because the mean-field, self-energy, and
exciton problems are treated on equal footing with atomistic detail, avoiding
any effective mass approximations.  However, compared to fully ab initio
approaches, ours is computationally efficient due to the tight-binding
parameterization and classical treatment of electrostatic screening.  The
results will be validated against recently obtained experimental data.  We
analyze the orbital character of the transitions, the dimensionality-dependent
nonhydrogenic Rydberg series of the excitons, and present a term-by-term
analysis of the carrier confinement and electrostatic effects that determine the
optical properties of the layered HOIPs.

\begin{figure}[t]
    \centering
    \includegraphics{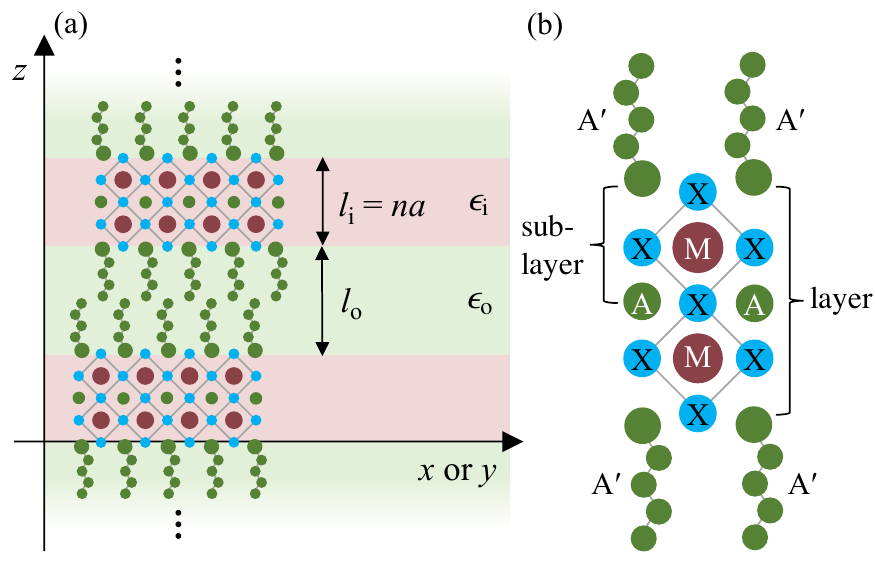}
    \caption{(a) The electrostatic model of a layered HOIP, with $n=2$ shown as
an example. For the case of
(C$_4$H$_9$NH$_3$)$_2$(CH$_3$NH$_3$)$_{n-1}$Pb$_n$I$_{3n+1}$, $a=6.39$~\AA
\cite{oku2015crystal}, $l_\mathrm{o}=8.81$~\AA \cite{muljarov1995excitons},
$\epsilon_\mathrm{i}=6.1$ \cite{ishihara1990optical}, and
$\epsilon_\mathrm{o}=2.1$ \cite{ishihara1990optical}. (b) A schematic
representation of the structure of the layered HOIP depicted in (a). A is a
small organic ammonium cation, A$^{'}$ is a bulky organic ammonium cation, M is
a metal cation, and X is a halogen anion.}
    \label{fig:model}
\end{figure}

As depicted in Fig.~\ref{fig:model}, we assume a layered HOIP
A$^{\prime}_2$A$_{n-1}$M$_n$X$_{3n+1}$, which has alternating layers composed of
organic spacers A$^{\prime 2+}_2$ and primarily-inorganic perovskites
A$_{n-1}$M$_n$X$_{3n+1}^{2-}$. The organic cations A and A$^\prime$ primarily
determine the structure and do not participate in the valence chemical
bonding~\cite{gao2016quasiparticle}.  Therefore, our tight binding model is
constructed based on the metal-halide perovskite structure M$_n$X$_{3n+1}$.
Furthermore, due to the long organic spacers, we neglect the hybridization
between perovskite layers~\cite{even2014understanding, stoumpos2017high}.  For
simplicity, we assume a cubic perovskite structure (space group $Pm$-$3m$), such
that M$_n$X$_{3n+1}$ is modeled as $n$ sublayers of the cubic MX$_3$ with an
additional X on the boundary.  This approximation precludes a study of the
Rashba effect~\cite{kim2014switchable, zheng2015rashba}, arising from the
inversion symmetry breaking of the typical orthorhombic
structure~\cite{stoumpos2016ruddlesden}, although this effect could be
straightforwardly included along the lines of
Ref.~\onlinecite{boyer2016symmetry}.  Henceforth, we will focus our attention on
the specific case of layered HOIPs made from the popular parent compounds of
ammonium or methylammonium (MA) lead iodide using butylammonium spacers,
i.e.~A$^{\prime+}$=C$_4$H$_9$NH$_3^+$, A$^{+}$=NH$_4^{+}$ or CH$_3$NH$_3^+$,
M=Pb, and X=I.

With the above approximations, we construct a tight-binding model in the basis
of the atomic orbitals corresponding to the Pb 6s and 6p orbitals and the I 5s
and 5p orbitals.  This approach has been successfully applied to the 3D HOIPs,
in order to study a variety of effects including the Rashba
splitting~\cite{kim2014switchable, zheng2015rashba}, the dependence of
electronic properties on the atomic structure \cite{knutson2005tuning,
boyer2016symmetry}, and dynamic properties~\cite{mayers2018lattice}.  The
parameters of our tight-binding model are determined from a single DFT
calculation on the cubic crystal structure; for simplicity, we perform
calculations on CsPbI$_3$, whose lattice constant is very similar to that of
cubic MAPbI$_3$ and NH$_4$PbI$_3$ (all between $6.2-6.3~\AA$).  DFT calculations
were performed using Quantum Espresso\cite{QE-2017} with the PBE exchange
correlation functional~\cite{perdew1996generalized} and tight-binding matrix
elements were determined using the Wannier90 code\cite{mostofi2014updated}.
From these calculations, a number of symmetry unique real-space matrix elements
(which extend beyond nearest-neighbor) can be determined and used to build a
tight-binding representation for any \textit{layered} HOIP structure.
Importantly, this construction captures the carrier confinement (kinetic energy)
effect arising in nanostructured materials.

Next, we include the spin-orbit coupling (SOC), which is extremely large because
of the high atomic numbers of Pb and I~\cite{even2013importance}.  Including SOC
requires two spin-orbital basis functions for each atomic spatial orbital
leading to eight spin-orbitals for each Pb atom and eight spin-orbitals for each
I atom.  Following Ref.~\onlinecite{kim2014switchable}, we add two on-site
Hamiltonians for the Pb and I p orbitals, each of the form
\begin{equation}
h^{\mathrm{SOC,B}}=\frac{\Delta_{\mathrm{SOC}}^\mathrm{B}}{3}
\mqty(  0 & -i & 0 & 0 & 0 & 1 \\
        i & 0 & 0 & 0 & 0 & -i \\
        0 & 0 & 0 & -1 & i & 0 \\
        0 & 0 & -1 & 0 & i & 0 \\
        0 & 0 & -i & -i & 0 & 0 \\
        1 & i & 0 & 0 & 0 & 0 ),
\end{equation}
where the basis is ordered as 
$\lbrace\ket*{p_x\uparrow}$, $\ket*{p_y\uparrow}$,
$\ket*{p_z\uparrow}$, $\ket*{p_x\downarrow}$, $\ket*{p_y\downarrow}$,
$\ket*{p_z\downarrow}\rbrace$ and where $\Delta_{\mathrm{SOC}}^{\mathrm{B}}$ 
is the spin-orbit coupling constant of atom type B.  We emphasize that we do not
treat SOC perturbatively because of the large magnitude and the resulting bands
have mixed spin character, i.e.~$S_z$ is not a good quantum number.

Combining the DFT and SOC matrix elements produces the total $\vk$-dependent
mean-field tight-binding Hamiltonian with matrix elements,
\begin{equation}
h^\mathrm{DFT+SOC}_{\lambda\mu}(\vk)
    = \sum_{\vT}e^{i\vk\cdot\vT} 
        h^\mathrm{DFT}_{\lambda\bm{0},\mu\bm{T}}
    + h^{\mathrm{SOC,Pb}}_{\lambda\bm{0},\mu\bm{0}} 
    + h^{\mathrm{SOC,I}}_{\lambda\bm{0},\mu\bm{0}}
\end{equation}
where $\lambda$ and $\mu$ are the atomic spin-orbital types and $\bm{T}$ is 
a lattice translation vector.  Diagonalization of this mean-field Hamiltonian
yields an approximate layered HOIP band structure with SOC included
nonperturbatively.  However, the DFT band gap is a poor approximation to the
true quasiparticle band gap, which can be estimated using the GW approximation
to the self-energy.


The GW approximation to the self-energy, $\Sigma = i GW$, requires the screened
Coulomb interaction $W$, which is typically calculated within the random-phase
approximation~\cite{hedin1965new,strinati_dynamical_1980,strinati_dynamical_1982,hybertsen1985first,hybertsen1986electron}.
Building $W$ in this manner is responsible for the cost of GW calculations.
Instead, we employ a few parameters that correct the 3D bulk band structure of
the parent HOIP and then we appeal to a classical electrostatic treatment of
screening in order to calculate the \textit{change} in the self-energy due to a
different dielectric environment (as realized in layered structures).  

\begin{figure}[b]
    \centering
    \includegraphics{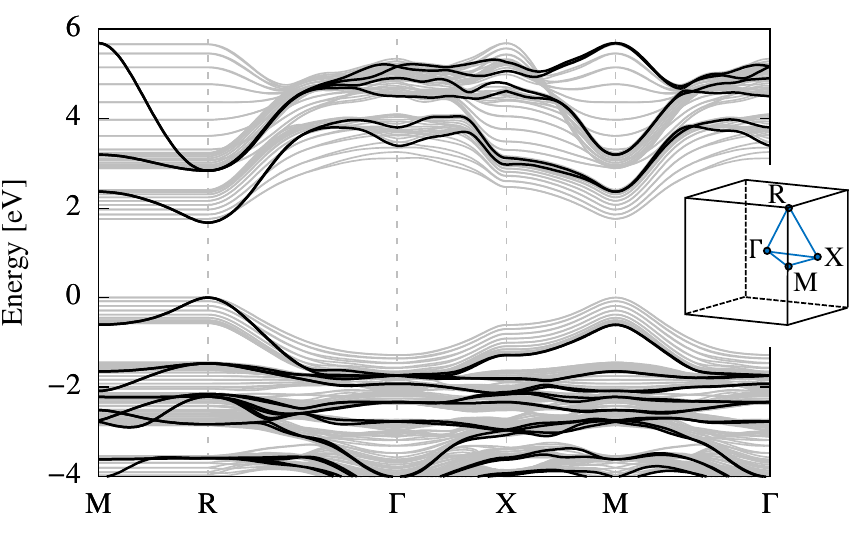}
    \caption{Band structure of the 3D HOIP calculated by the tight binding
method including the spin-orbit coupling and the GW self-energy correction using
Eq.~\ref{eq:bulk}.  Grey curves show the band structure of the layered HOIP with
$n=10$ calculated using Eq.~\ref{eq:TB_n}. The cubic Brillouin zone is shown
along with high symmetry points and $k$-point path; note that in the $n=\infty$ supercell
limit, the M and R points become equivalent.}
    \label{fig:bulk}
\end{figure}

First, for the bulk correction, we identify five symmetry-distinct orbital types,
corresponding to the Pb s orbitals, the Pb p orbitals (all three equivalent),
the I s orbitals, the I p orbitals that are perpendicular to the Pb-I-Pb bond
and the I p orbitals that are parallel to the Pb-I-Pb bond.  For each
of these atom types, we introduce a constant self-energy $\Sigma_\mu$,
all of which are added to the diagonal of the above DFT+SOC Hamiltonian, in order
to create an approximate GW Hamiltonian for the \textit{bulk} HOIP,
\begin{equation}
h^{\mathrm{GW}}_{\lambda\mu}(\vk) =
h^\mathrm{DFT+SOC}_{\lambda\mu}(\vk) 
    + \Sigma_{\lambda}\delta_{\lambda\mu}
\label{eq:bulk}
\end{equation}
with band energies $\varepsilon(\vk)$ and eigenvectors $\mathbf{C}(\vk)$
determined by
\begin{equation}
\mathbf{h}^{\mathrm{GW}}(\vk)\mathbf{C}(\vk) = \mathbf{C}(\vk)\varepsilon(\vk).
\end{equation}
These five self-energies $\Sigma_\mu$, the above two SOC constants
$\Delta_{\mathrm{SOC}}^{\mathrm{B}}$, and a single rigid ``scissors''-style
shift of the conduction band $\Sigma_\mathrm{CB}$ are then optimized in order to
best reproduce the \textit{ab initio} GW band structure of MAPbI$_3$ presented
in Ref.~\onlinecite{brivio2014relativistic}, which also included SOC.
The resulting band structure is shown in Fig.~\ref{fig:bulk} with a band gap of
1.67~eV and an effective mass of $0.20m_0$ for both the electron and the hole.
The excitonic reduced mass of $0.10m_0$ is consistent with the value determined
by magneto-absorption experiments, $0.104m_0$~\cite{miyata2015direct}. 
The spin-orbit
coupling constants are determined to be 
$\Delta_{\mathrm{SOC}}^{\mathrm{Pb}}=1.18$ eV and
$\Delta_{\mathrm{SOC}}^{\mathrm{I}}=1.06$ eV, 
which are similar to the values used in a previous tight-binding 
study~\cite{boyer2016symmetry}.

Moving on to layered HOIPs, we model the system as a stack of
alternating organic and inorganic slabs, each with a uniform dielectric constant
as shown in Fig. \ref{fig:model}.  The screened Coulomb interaction between two
charges placed on the planes at
$z_1$ and $z_2$ and separated by the in-plane coordinate $\rho$, 
$W(\rho, z_1, z_2)$, can be obtained in a
closed form when the number of layers is
infinite~\cite{guseinov1984coulomb,muljarov1995excitons} or via the
electrostatic transfer matrix method~\cite{cavalcante2018electrostatics} 
when the number of layers is finite. 
Relative to the bulk self-energy, the \textit{change} in the electrostatic 
self-energy of a charge in the inorganic section of a layered HOIP at height $z$ is given 
by~\cite{brus1983simple, kumagai1989excitonic}
\begin{equation}
\delta\Sigma(z)=\frac{1}{2}\lim_{\rho\to 0} \qty[W(\rho, z_1=z_2=z)-\frac{1}{\epsilon_i\:\rho}],
\end{equation}
where $\epsilon_i$ is the dielectric constant of the inorganic section.

In the spirit of the conventional one-shot application of the GW approximation,
we correct our bulk band structure using $\delta\Sigma(z)$ via perturbation 
theory.  Specifically, we assign the $z$-dependent self-energy shift to all
atoms $\mu$ at the height $z_\mu$ and for each band we correct the eigenvalues
with an
expectation value, 
\begin{subequations}
\label{eq:TB_n}
\begin{align}
E_c(\vk) &= \varepsilon_c(\vk) + \delta\Sigma_c(\vk) 
    = \varepsilon_c(\vk) +\sum_{\mu} \left|C_{\mu c}(\vk)\right|^2
        \delta\Sigma(z_\mu) \\
E_v(\vk) &= \varepsilon_v(\vk) + \delta\Sigma_v(\vk)
    = \varepsilon_v(\vk) -\sum_{\mu} \left|C_{\mu v}(\vk)\right|^2
        \delta\Sigma(z_\mu)
\end{align}
\end{subequations}
where $c$ and $v$ index the spin-orbital conduction and valence bands
(which have mixed spin character).
In Fig.~\ref{fig:bulk}, we compare the bulk band structure
to that obtained for the layered HOIP with $n=10$.  As expected,
the latter has about the same band gap, but a ten-fold band multiplicity
that tracks the dispersion of the bulk band structure.

To simulate the optical spectrum, we solve the Bethe-Salpeter equation (BSE)
based on the GW
self-energy~\cite{sham_many-particle_1966,hanke_many-particle_1980,strinati_effects_1984,albrecht_excitonic_1998,rohlfing2000electron,
ridolfi2018excitonic}.  The BSE, in the Tamm-Dancoff approximation, follows from
the wavefunction ansatz
\begin{equation}
|\Psi_X\rangle = \sum_{cv} \sum_\vk A_{cv}^X(\vk) 
    a_{c,\vk}^\dagger a_{v,\vk} |0\rangle,
\end{equation}
where the exciton wavefunction $A_{cv}^X(\vk)$ and excitation energy
$E_X$ are determined by the eigenvalue equation
\begin{equation}
\begin{split}
E_X A^X_{cv}(\vk)
&= \left[E_c(\vk)-E_v(\vk)\right]A_{cv}^X(\vk) \\
&\hspace{1em} - \sum_{c^\prime v^\prime} \sum_{\vk^\prime} 
        (c\vk,c^\prime\vk^\prime|v^\prime\vk^\prime v\vk)_W
            A^X_{c^\prime v^\prime}(\vk^\prime),
\end{split}
\end{equation}
where we have neglected the excitonic exchange interaction.
The two-electron integral associated with the screened Coulomb interaction
is
\begin{equation}
\begin{split}
(c\vk,c^\prime\vk^\prime|v^\prime\vk^\prime v\vk)_W
    &= \sum_{\lambda\mu}
        C_{\lambda c}^*(\vk)
        C_{\mu c^\prime}^*(\vk^\prime)
        C_{\mu v}(\vk)
        C_{\lambda v^\prime}(\vk^\prime) \\
    &\hspace{3em} \times W(|\vk-\vk^\prime|, z_\lambda, z_\mu).
\end{split}
\end{equation}
Here, $W(k,z_1,z_2)$ is the in-plane Fourier-transformed screened Coulomb
interaction obtained from the classical electrostatic problem described
in the previous section.  Furthermore, we have approximated the two-electron
integrals in the localized atomic-orbital basis by neglecting
differential overlap, 
$\phi^*_\lambda(\vr)\phi_\mu(\vr)d\vr 
\approx \delta_{\lambda\mu}|\phi_\mu(\vr)|^2 d\vr$.

Based on the solutions of the BSE, we calculate the linear absorption via
the usual Fermi's golden rule, leading to
\begin{equation}
I(\omega) \propto
    \sum_X \bigg|\sum_{cv}\sum_\vk A_{cv}^X(\vk)
        \bm{\lambda}\cdot\vp_{cv}(\vk)\bigg|^2 
        \delta(\hbar\omega - E_X),
\end{equation}
where $\bm{\lambda}$ is the polarization of the light
and $\vp_{cv}$ are transition matrix elements of the momentum 
operator. The momentum matrix
elements are obtained via the commutation relation
$\vp(\vk) =(-im/\hbar)[\vr, H]$ as follows~\cite{pedersen2001optical}
\begin{equation}
\begin{split}
\vp_{cv}(\vk) &= \frac{m}{\hbar} 
    \left[ \nabla_\vk \mathbf{h}^{\mathrm{GW}}(\vk) \right]_{cv} \\
    &\hspace{1em} + i\frac{m}{\hbar} \left[E_c(\vk)-E_v(\vk)\right]
        \sum_{\lambda\mu} \vr_{\lambda\mu}
        C^*_{\lambda c}(\vk) C_{\mu v}(\vk)
\label{Eq:mom}
\end{split}
\end{equation}
where the tight-binding dipole matrix elements $\vr_{\lambda\mu}$ are also
obtained from Wannier90.  We emphasize that this approach includes both inter-
and intra-atomic transitions.

All BSE results shown are calculated on an $N_k\times N_k$ Monkhorst-Pack mesh
with $N_k=50-70$, for which the first few excitation energies are
well-converged.  The BSE Hamiltonian is constructed using two valence bands and
two conduction bands.  In our testing, this approximation introduces an error of
less than 0.01~eV in the absolute value of the excitation energies.
Importantly, our calculations account for the band structure throughout the
Brillouin zone, which means that nonparabolicity is explicitly treated and the
calculation is more realistic than the use of an effective mass model.


\begin{figure}
    \centering
    \includegraphics{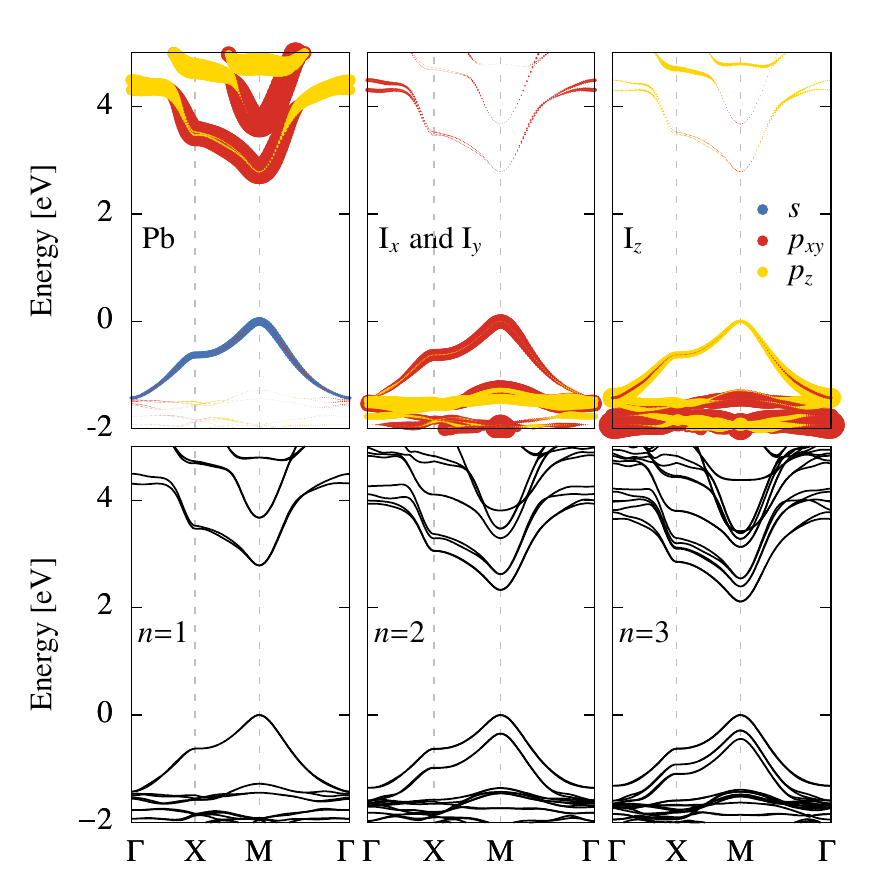}
    \caption{Band structures calculated by Eq.~\ref{eq:TB_n}. The top row shows
orbital-projected band structures of $n=1$ for each atom and orbital type, where
the line width is proportional to the contribution.  The red lines for $p_{xy}$
are the sum of the $p_x$ and $p_y$ orbital contributions. The bottom row shows
the band structures of $n=1$, 2, and 3.}
    \label{fig:ndep}
\end{figure}

There are three types of atoms in the inorganic layer: Pb, horizontal I (I$_x$
and I$_y$), and vertical I (I$_z$). Figure \ref{fig:ndep} shows the contribution
of each atom type's orbitals to the band structure of $n=1$. At the M point, the
valence band maximum (VBM) has 53\% I$_{xy}$ $p$ orbital character and 32\% Pb
$s$ orbital character. The contribution of I$_z$ shows a clear difference from
that of I$_{xy}$, which is only 15\% at the M point and increases to 86\% at
$\mathrm{\Gamma}$ point. The conduction bands are dominated by Pb, with the
conduction band minimum (CBM) having 95\% Pb $p$ orbital character.  For this
reason, the spin-orbit splitting at the CBM is almost entirely determined by the
Pb $p$ orbitals.
These VB and CB orbital compositions are in agreement with previous DFT
studies~\cite{bala2018atomic, molina2018excitonic}.  As we go to larger $n$, the
bands multiply and split due to the strong coupling between the sublayer. As
shown in Fig.~\ref{fig:ndep}, each band from $n=1$ splits into $n$ subbands for
larger $n$, which leads to a reduction in the band gap. For large $n$, the band
structure of the bulk HOIP is regained, as demonstrated in Fig.~\ref{fig:bulk}.

\begin{figure}[t]
    \centering
    \includegraphics{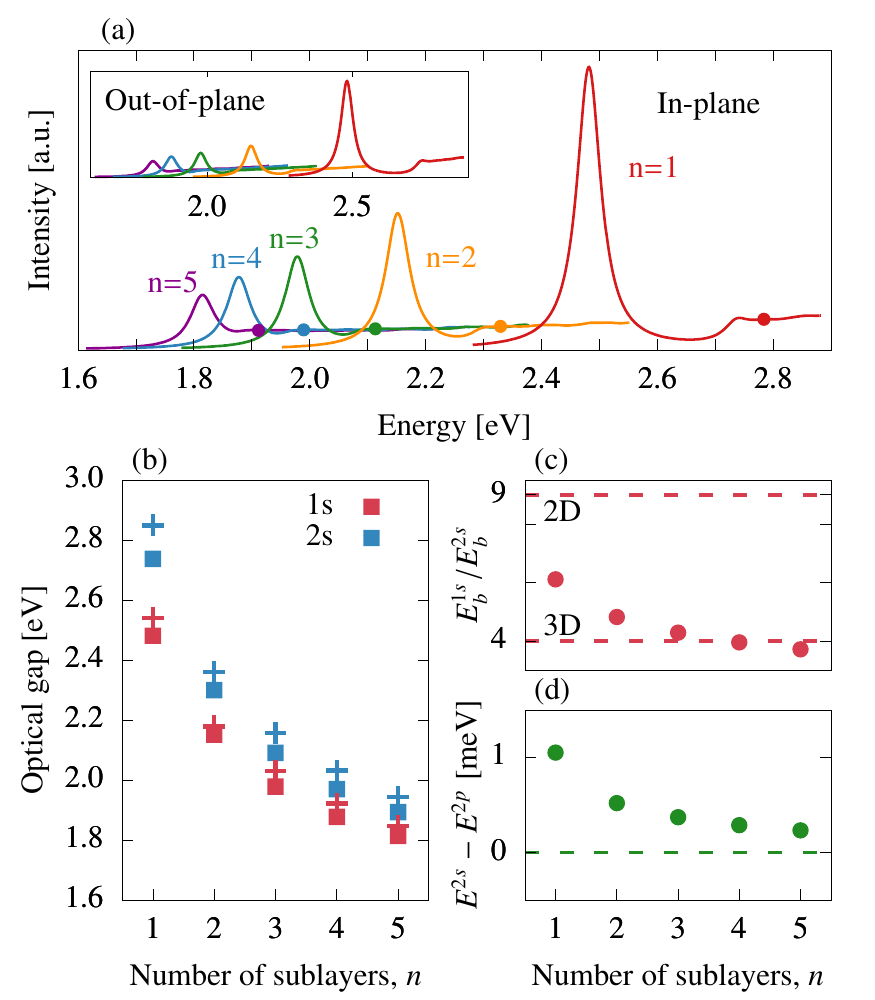}
    \caption{(a) Excitonic absorption spectra of $n=1$ through 5 for in-plane
polarized light. Solid dots indicate the band gap of each system.  The inset
shows the spectra for out-of-plane polarized light, with a magnified scaled
compared to the main figure. The spectra are plotted using a Lorentzian line
shape with a broadening of 50 meV (full width at half maximum).  (b) 1s and 2s
peak position as a function of $n$. Plus symbols ($+$) are the experimentally
measured values from Ref.~\onlinecite{blancon2018scaling}.  (c) Ratio of the 1s
and 2s exciton binding energies, where a ratio of 9 or 4 is the prediction of an
ideal 2D or 3D hydrogenic model of excitons. (d) Energy difference between the
2s and 2p excitons, showing the degeneracy breaking at small $n$.} 
    \label{fig:spec_opt}
\end{figure}

The calculated absorption spectra shown in Fig.~\ref{fig:spec_opt}(a) exhibit a
strong 1s peak and weak 2s and 3s peaks; the quasiparticle band gap is indicated
with a solid circle.  As $n$ increases, both the absorption energy and intensity
decrease, approaching the bulk spectrum.  For the out-of-plane polarization, the
first term of Eq.~\ref{Eq:mom} is zero, and absorption occurs purely due to the
intra-atomic dipole moment.  Both polarizations gives a spectrum of similar
shape, but the out-of-plane polarization has a much smaller absorption
intensity, about 5\% that of the in-plane polarization.  The energies of the 1s
and 2s states are in good agreement with recent experimental measurements
\cite{blancon2018scaling}, as shown in Fig.~\ref{fig:spec_opt}(b); the
discrepancy is less than 0.1~eV for both states and for all values of $n$.  In
our calculations, the 1s exciton binding energies for $n=1-5$ are found to be
302~meV, 177~meV, 135~meV, 112~mev, and 97~meV.

Excitons in semiconductors are commonly analyzed with a hydrogenic model of the
interacting electron and hole.  In three dimensions, this model leads to s-type
exciton energy levels with binding energies $E_\mathrm{b}(m\mathrm{s}) =
\mathrm{Ry}/m^2$, where the Rydberg constant is $\mathrm{Ry} =
\mu/(2\varepsilon^2)$, $\mu$ is the exciton reduced mass, and $\varepsilon$ is a
uniform dielectric constant.  We can use this model to estimate the prediction
of our model for $n=\infty$, corresponding to the bulk HOIP.  Using $\mu =
0.10m_0$ (consistent with the band masses reported above) and $\varepsilon =
6.1$, we find $E_\mathrm{b}^\mathrm{1s} = 37$~meV.  This is in reasonable
agreement with experimental values, which range from 7.4 to
50~meV~\cite{ziffer2016electroabsorption,phuong2016free,yang2015comparison,miyata2015direct,hirasawa1994magnetoabsorption,ishihara1990optical,tanaka2003comparative}. 

Nonhydrogenic behavior in the exciton series~\cite{chernikov2014exciton} can be
caused by the combination of nonparabolicity in the band structure,
finite-thickness effects, and inhomogeneous dielectric screening.  Because the
layered HOIPs have some two-dimensional character, we will also compare to the
prediction of the 2D hydrogen model of excitons, $E_\mathrm{b}^{m\mathrm{s}} =
\mathrm{Ry}/(m-1/2)^2$.
The difference between the 2D exciton series and the 3D exciton series can be
seen in the ratio of the binding energies of the 1s and 2s states: in 2D
$E_\mathrm{b}^{1\mathrm{s}}/E_\mathrm{b}^{2\mathrm{s}} = 4$ and in 3D
$E_\mathrm{b}^{1\mathrm{s}}/E_\mathrm{b}^{2\mathrm{s}} = 9$.  In
Fig.~\ref{fig:spec_opt}(c), we show this ratio as a function of $n$.  At $n=5$,
the ratio is approximately 4, indicating a conventional 3D exciton within the
hydrogenic model.  At $n=1$, the ratio is about 6, i.e.~larger than 4 but
smaller than 9.  Therefore, the $n=1$ excitons are intermediate between 2D and
3D, displaying a nonhydrogenic Rydberg series.  This deviation from the ideal 2D
limit of 9 is due to the form of the screened Coulomb potential: the
heterogeneous dielectric yields an interaction that is not simply of the form
$1/\varepsilon r$.

Another manifestation of the nonuniform dielectric screening is the breaking of
accidental degeneracies in the exciton spectrum.  In particular, the angular
momentum degeneracy is broken for alternative interactions.  In
Fig.~\ref{fig:spec_opt}(d), we plot the energy difference
$E^{\mathrm{2s}}-E^{\mathrm{2p}}$ as a function of $n$.  At large $n$, the 2s
and 2p states are nearly degenerate, again indicating conventional 3D hydrogenic
exciton behavior.  However, the degeneracy is broken at small $n$; in
particular, at $n=1$ the 2p state is lower in energy by about 1~meV.

\begin{figure}
    \centering
    \includegraphics{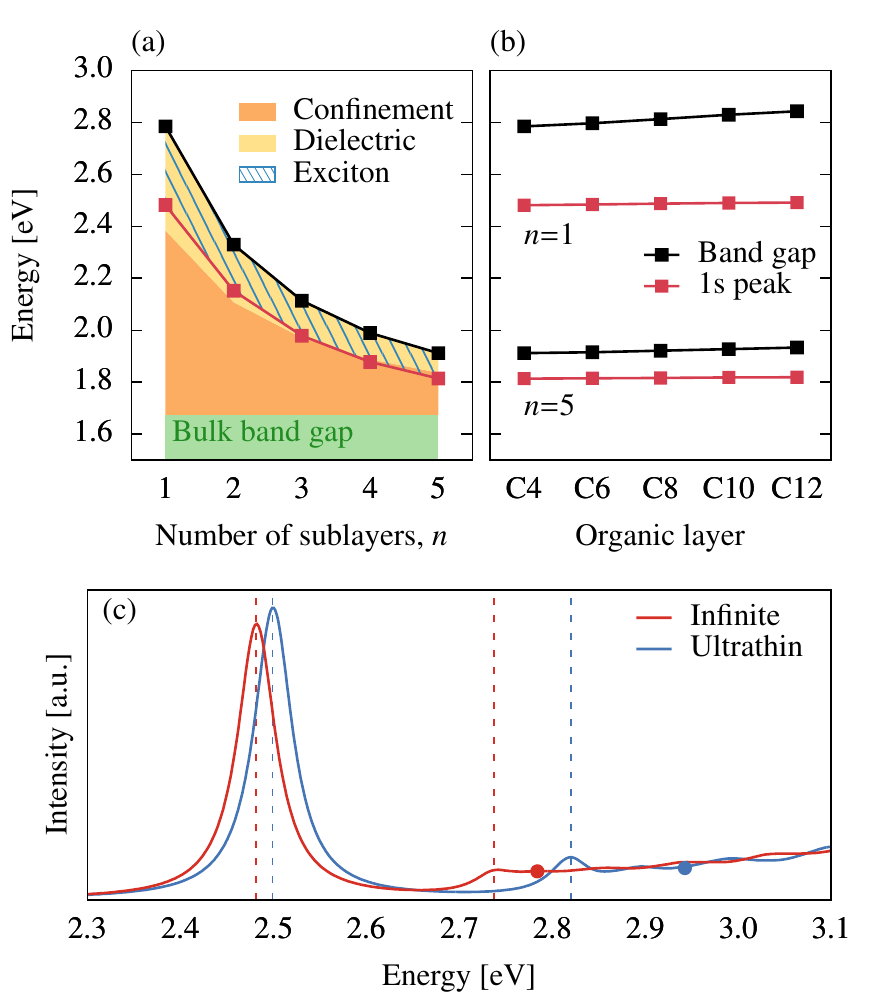}
    \caption{(a) Band gap (black) and optical gap (red) of the layered HOIP as a
function of $n$. Orange and yellow shaded area indicate the contribution of the
carrier confinement and dielectric contrast self-energy effects to the band gap.
Blue striped pattern shows the reduction of the optical gap due to the exciton
binding energy. The bulk band gap from Figure~\ref{fig:bulk} is shown as a green
shaded area. (b) Band gap and optical gap of the HOIP with $n=1$ and 5 as a
function of the thickness of the organic layer. C$m$ on the x axis indicates the
organic cation A$^{'+}$=C$_m$H$_{2m+1}$NH$_3^+$. (c) Absorption spectra of an
infinitely thick $n=1$ sample (red line) and an ultrathin exfoliated $n=1$
bilayer (blue line).  Solid dots are the position of the band gap. Dashed
vertical lines indicate the position of the 1s and 2s excitons.}
    \label{fig:conf_thin}
\end{figure}

Our calculation allows us to quantify the physical mechanisms responsible for
the thickness-dependent excitation energies, shown in
Fig.~\ref{fig:conf_thin}(a).  The $n$-dependent band gap, shown with black
symbols, can be decomposed into a bulk band gap of 1.67~eV, a carrier
confinement energy, and a self-interaction energy due to dielectric
contrast~\cite{katan2019quantum}.  The latter two effects make a significant
contribution to the band gap for $n=1-5$.  At $n=1$, the carrier confinement and
dielectric contrast increase the band gap by about 0.7~eV and 0.4~eV,
respectively.  By $n=5$, the contributions are more modest and amount to about
0.1~eV each.  A similar analysis can be done for the 1s excitation energy, which
has an additional negative contribution arising from the exciton binding energy.
The exciton binding energy almost exactly cancels the band gap increase due to
dielectric contrast, which is analogous to the insensitivity of the first
excitation energy of transition-metal dichalcogenides to the dielectric of the
environment~\cite{cho2018environmentally}.  Therefore, the thickness-dependence
of the first excitation energy is largely determined by the confinement energy.
However, the wavefunction character of the excited state is excitonic and
distinct from that of a confined but noninteracting electron-hole pair.  This
wavefunction difference will impact various properties and is responsible for
the strong, narrow peak observed in the absorption spectrum.

The above result suggests that the band gap can be tuned by the organic spacer
but that the 1s excitation energy will be relatively insensitive, assuming that
the structure of the inorganic layer is preserved.  In
Fig.~\ref{fig:conf_thin}(b), we show the band gap and 1s peak energy as a
function of the length of the organic molecules used as spacers, ranging from
butylammonium (C4) to dodecylammonium (C12).  Indeed, for $n=1$, the band gap
increases by about 0.1~eV by increasing the length of the organic cation, but
the 1s energy is essentially unchanged, consistent with early experimental
observations~\cite{ishihara1990optical}.  The effect is less dramatic for $n=5$,
due to the decreased relevance of the environment surrounding a thick inorganic
layer.

An alternative approach to tune the environment of the inorganic layer is
through mechanical exfoliation, like can be achieved in layered van der Waals
materials and was pursued for HOIPs in Ref.~\onlinecite{yaffe2015excitons}.  In
Fig.~\ref{fig:conf_thin}(c), we compare the spectrum of the $n=1$ bulk crystal
and an exfoliated $n=1$ bilayer on a SiO$_2$/Si substrate (modeled with a
dielectric constant $\varepsilon=2.13$). This latter ultrathin sample was the
one isolated and studied in Ref.~\onlinecite{yaffe2015excitons}.  The 1s peak of
the ultrathin bilayer is located at 2.50~eV, only 20~meV higher than the case of
the infinite number of alternating layers and in good agreement with the
experimental value of 2.53~eV. The 2s exciton state, having a smaller binding
energy, exhibits a larger increase by 90~meV. The band gap increase of 160~meV
shows the major consequence of the change in the degree of dielectric contrast
with a concomitant increase in the exciton binding energy from 302~meV (bulk) to
444~meV (ultrathin).  This latter value is in good agreement with the
experimentally inferred value of 490~meV.

In conclusion, a semiempirical tight-binding model of the layered HOIPs has been
developed and parameterized by ab initio DFT and GW calculations of the 3D HOIP.
Our model, together with an approximate solution of the Bethe-Salpeter equation
for neutral excitons, provides a tool to analyze the electronic band structure,
absorption spectrum, and exciton properties at a low computational cost compared
to fully ab initio calculations, which are daunting due to the large unit cell
size of the layered HOIPs.  

Our results are in good agreement with recently reported experimental
measurements.  Most importantly, the level of theory enables direct access to
the physical properties underlying the observed behaviors.  In particular, the
carrier confinement, dielectric self-energy, and exciton binding energies have
been systematically studied and quantified as a function of $n$.  We have
demonstrated that the 1s and 2s excitons exhibit a nonhydrogenic behavior --
with respect to level dispositions and angular momentum degeneracy -- that can
be controlled by $n$ and that conventional hydrogenic behavior is recovered at
large $n$. 

This general approach is promising for the investigation of other layered
compounds with complex unit cells.  Furthermore, the microscopic but affordable
tight-binding description will enable the treatment of additional phenomena in
the layered HOIPs, which are challenging at the ab initio level.  Specifically,
we anticipate the study of noncubic phases~\cite{menendez2014self,
stoumpos2016ruddlesden} and the Rashba effect~\cite{kim2014switchable,
zheng2015rashba}, lead ion displacements and other sources of
disorder~\cite{even2014analysis, motta2016effects, smith2017decreasing}, and
electron-phonon coupling~\cite{wu2015trap, guo2016electron}, which can impact
screening and renormalize the exciton binding energy.  Finally, the large
exciton binding energies found here suggest the presence of more exotic
multi-carrier complexes such as trions~\cite{zahra2019screened} and
biexcitons~\cite{kato2003extremely, elkins2017biexciton, thouin2018stable}; work
along these lines is currently in progress.
 
\begin{acknowledgments}
We thank Shi-Ning Sun for early work on this project. 
Calculations were performed using resources provided by the University of
Chicago Research Computing Center and the Flatiron Institute.
This work was supported by the US-Israel Binational Science Foundation Grant
BSF-2016362 and by the Air Force Office of Scientific Research under award
number FA9550-18-1-0058. The Flatiron Institute is a division of the Simons
Foundation.
\end{acknowledgments}

\providecommand{\latin}[1]{#1}
\makeatletter
\providecommand{\doi}
  {\begingroup\let\do\@makeother\dospecials
  \catcode`\{=1 \catcode`\}=2 \doi@aux}
\providecommand{\doi@aux}[1]{\endgroup\texttt{#1}}
\makeatother
\providecommand*\mcitethebibliography{\thebibliography}
\csname @ifundefined\endcsname{endmcitethebibliography}
  {\let\endmcitethebibliography\endthebibliography}{}

\end{document}